\documentclass[a4paper,11pt]{article}
\usepackage{latexsym}
\usepackage{feynmf}		
\def\ba{\begin{eqnarray}}
\def\ea{\end{eqnarray}}

\def\lb{\label}
\def\be{\begin{equation}}
\def\ee{\end{equation}}

\begin{document}
\baselineskip0.25in
\title{Gauss-Bonnet models with cosmological constant and non zero spatial curvature in  $D=4$}
 \author{Juan Manuel Armaleo \thanks{Departamento de F\'isica, UBA, Buenos Aires, Argentina juanma.armaleo@gmail.com}, Juliana Osorio Morales \thanks{Departamento de Matem\'aticas Luis Santal\'o (IMAS), UBA CONICET, Buenos Aires, Argentina
juli.osorio@gmail.com.} and Osvaldo P. Santill\'an \thanks{Departamento de Matem\'aticas Luis Santal\'o (IMAS), UBA CONICET, Buenos Aires, Argentina
firenzecita@hotmail.com and osantil@dm.uba.ar.}}

\date {}
\maketitle

\begin{abstract}
In the present paper the possibility of eternal universes in Gauss-Bonnet theories of gravity in four dimensions is analysed.
It is shown that, for zero spatial curvature and zero cosmological constant, if the coupling is such that $0<f'(\phi)\leq c \exp(\frac{\sqrt{8}}{\sqrt{10}}\phi)$, then there are solutions that are eternal. Similar conclusions are found when a cosmological constant turned on. These conclusions are not generalized for the case when the spatial curvature is present, but we are able to find
some general results about the possible nature of the singularities. The presented results correct some dubious arguments in \cite{yogo}, although the same conclusions are reached. On the other hand, these past results are considerably generalized to a wide class of situations which were not considered in \cite{yogo}.

\end{abstract}

\section{Introduction}

One of the main interests in higher derivative gravity theories is that they can describe inflation
by the addition of a higher order curvature to the Einstein-Hilbert action \cite{capo1}-\cite{capo2}.
This is achieved without the addition of dark energy or scalar fields. An important role in this context is played by the Gauss-Bonnet invariant, since it appears in QFT renormalization in curved space times
\cite{birrell}. In addition, the Gauss-Bonnet term arises in low-energy effective actions of some string theories.
For instance, the tree-level string effective action has been calculated up to several orders in
the $\alpha'$ expansion  in \cite{3}-\cite{4}. The result is that 
there is no moduli dependence of the tree-level couplings. However, one loop corrections
to the gravitational couplings have been considered in the context of
orbifold compactifications of the heterotic superstring \cite{5}. It has been shown in that reference that
there are no moduli dependent corrections to the Einstein term while there are non trivial
curvature contributions. They appear as the Gauss-Bonnet combination multiplied
by a function of the modulus field. 

The results described above partially motivated the study of cosmological consequences of the 
Gauss-Bonnet term. In four dimensions, this term does not
have any dynamical effect. However, when this term is non-minimally coupled
with any other field such as a scalar field $\phi$, the resulting dynamics is non trivial.
Several cosmological consequences has been exploited in recent literature, and we refer the reader
to \cite{s1}-\cite{Granda2} and references therein. But the aim of the present letter is not focused in inflationary
aspects of the theory, instead in the characterization of singular and eternal solutions of the theory.
It is important to mention that there exist preliminary works on this subject, examples are given in \cite{singularityfree}-\cite{singularityfree08}.
In particular, the results of \cite{singularityfree}-\cite{singularityfree2} suggest the existence of singular solutions as well as regular solutions. 
The singular solutions are confined to an small portion of the phase space, while the non singular fill the rest. This situation
is different than in GR, where the Gauss-Bonnet term is absent, and the powerful Hawking-Penrose theorems apply \cite{hp}.

In the present work we are going to provide evidence for these claims, when a cosmological constant is turned on or when the spatial curvature is vanishing. In addition, some partial results about the case with $k=\pm 1$ will be also presented.

This work is organized as follows. In section 2, generalities of Gauss-Bonnet models are briefly reviewed. Section 3 reviews some general arguments given in \cite{yogo} for zero spatial curvature and vanishing cosmological constant. These arguments are valid for any model of these characteristics. Section 4 contains an analysis of eternal universes which avoid some dubious arguments in \cite{yogo}. The calculations presented in this section are particularly explicit, since they are an important part of this paper. In section 5 the results of section 4 are generalized to the case where the cosmological constant is turned on. In section 6, the results of section 3 are generalized for the case in which the scalar curvature $k$ is turned on. The obtained results are not as universal as the ones in section 3, but some partial conclusions concerning the possible type of singularities can be obtained. In section 7 the results of section 4 are partially generalized to the case where the spatial curvature is turned on. Section 8 contains a discussion of the obtained results and open perspectives.

\section{Gauss-Bonnet equation}

The model that will be considered here is a generalization of the Gauss-Bonnet one. Recall that a pure Gauss-Bonnet gravity model is described in D-dimensions by the following action 
\begin{equation}
S_p=\int d^{D}x \sqrt{-g}G,
\label{accionGB}
\end{equation}
with $G$ being the Gauss-Bonnet invariant 
\begin{equation}
G\equiv R^2-4R_{\alpha \beta}R^{\alpha \beta}+R_{\alpha \beta\gamma\delta}R^{\alpha \beta\gamma\delta}.
\label{gaussbonnet1}
\end{equation}
The signature to be employed in the following is $(-, +, ..., +)$.
The equations of motions $\delta S=0$ that arise by minimizing the action with respect to variations $\delta g_{\mu\nu}$ are the following
$$
-\frac{1}{2}R_{\rho\sigma}R^{\rho\sigma}g_{\alpha\beta}-\nabla_{\alpha}\nabla_{\beta}R-2R_{\rho\beta\alpha\sigma}R^{\sigma\rho}+\frac{1}{2}g_{\alpha\beta}\Box R
+ \Box R_{\alpha\beta}+\frac{1}{2}R^{2}g_{\alpha\beta}-2RR_{\alpha\beta}
$$
\be
-2\nabla_{\beta}\nabla_{\alpha}R+2g_{\alpha\beta}\Box R -\frac{1}{2}R+R_{\alpha\beta}=0.
\label{gb}
\end{equation}
In four dimensions, the term $\sqrt{-g}G$ can be expressed as a total derivative
\begin{equation}
\sqrt{-g}G=\partial_{\alpha}K^\alpha,
\qquad 
K^\alpha=\sqrt{-g}\epsilon^{\alpha\beta\gamma\delta}\epsilon_{\rho\sigma}^{\mu\nu}\Gamma^{\rho}_{\mu\beta}\bigg[\frac{R^{\sigma}_{\nu\gamma\delta}}{2}+\frac{\Gamma^{\sigma}_{\lambda\gamma}\Gamma^{\lambda}_{\nu\delta}}{3}\bigg].
\label{lalala}
\end{equation}
Therefore in $D=4$ and with a manifold without boundary, this model is irrelevant.
However, the following modified action
\begin{equation}\label{g2}
S_m=\int d^4x \sqrt{-g}\left\{ \frac{1}{2\kappa^2}R
    - \frac{1}{2}\partial_\mu \phi \partial^\mu \phi+V(\phi)+ f(\phi) G\right\},
\end{equation}
is non trivial from the physical point of view, as the Gauss-Bonnet is coupled to the real scalar field $\phi$ by the coupling $f(\phi)$. Due to this coupling, this modified lagrangian it is not a total derivative and contributes to the equations of motion. 

In the following, a Gauss-Bonnet model with a potential $V(\phi)$ will be considered, as in references \cite{kanti1}-\cite{kanti2}. The equation for the scalar field $\phi$ in this case is given by
\begin{equation}
\label{g3}
\nabla^2 \phi+ f'(\phi) G-V'(\phi)=0.
\end{equation}
The equations of motion for the metric $g_{\mu\nu}$ are more involved. The variation of the action throws the following result
$$
0=\frac{1}{\kappa^2}\left(- R^{\mu\nu} + \frac{1}{2} g^{\mu\nu} R\right)+\frac{1}{2}\partial^\mu \phi \partial^\nu \phi
 - \frac{1}{4}g^{\mu\nu} \partial_\rho \phi \partial^\rho \phi+ \frac{1}{2}g^{\mu\nu}f(\phi) G +V(\phi)
-2 f(\phi) R R^{\mu\nu} 
$$
$$
+ 2 \nabla^\mu \nabla^\nu \left(f(\phi)R\right)- 2 g^{\mu\nu}\nabla^2\left(f(\phi)R\right) 
+ 8f(\phi)R^\mu_{\ \rho} R^{\nu\rho}- 4 \nabla_\rho \nabla^\mu \left(f(\phi)R^{\nu\rho}\right)
 - 4 \nabla_\rho \nabla^\nu \left(f(\phi)R^{\mu\rho}\right)
 $$
 $$
 + 4 \nabla^2 \left( f(\phi) R^{\mu\nu}  \right)+ 4g^{\mu\nu} \nabla_{\rho} \nabla_\sigma \left(f(\phi) R^{\rho\sigma} \right)
- 2 f(\phi) R^{\mu\rho\sigma\tau}R^\nu_{\ \rho\sigma\tau}+ 4 \nabla_\rho \nabla_\sigma \left(f(\phi) R^{\mu\rho\sigma\nu}\right).
$$
However, by taking into account the following identities
$$
\nabla^\rho R_{\rho\tau\mu\nu}= \nabla_\mu R_{\nu\tau} - \nabla_\nu
R_{\mu\tau},
$$
$$
\nabla^\rho R_{\rho\mu} = \frac{1}{2} \nabla_\mu R,
$$
$$
\nabla_\rho \nabla_\sigma R^{\mu\rho\nu\sigma} =
\nabla^2 R^{\mu\nu} - {1 \over 2}\nabla^\mu \nabla^\nu R
+ R^{\mu\rho\nu\sigma} R_{\rho\sigma}- R^\mu_{\ \rho} R^{\nu\rho},
$$
$$      
\nabla_\rho \nabla^\mu R^{\rho\nu}
+ \nabla_\rho \nabla^\nu R^{\rho\mu}
= {1 \over 2} \left(\nabla^\mu \nabla^\nu R
+ \nabla^\nu \nabla^\mu R\right)
 - 2 R^{\mu\rho\nu\sigma} R_{\rho\sigma}
+ 2 R^\mu_{\ \rho} R^{\nu\rho},
$$
$$
\nabla_\rho \nabla_\sigma R^{\rho\sigma} = \frac{1}{2} \Box R,
$$
which are consequences of the Bianchi identities, the last expression can be written as \cite{Nojiri}
$$
0=\frac{1}{\kappa^2}\left(- R^{\mu\nu} + \frac{1}{2} g^{\mu\nu} R\right)
      +  \left(\frac{1}{2}\partial^\mu \phi \partial^\nu \phi
      - \frac{1}{4}g^{\mu\nu} \partial_\rho \phi \partial^\rho \phi+V(\phi) \right)
   + \frac{1}{2}g^{\mu\nu}f(\phi) G
$$   $$  
   -2 f(\phi) R R^{\mu\nu} + 4f(\phi)R^\mu_{\ \rho} R^{\nu\rho}
   -2 f(\phi) R^{\mu\rho\sigma\tau}R^\nu_{\ \rho\sigma\tau}
+4 f(\phi) R^{\mu\rho\sigma\nu}R_{\rho\sigma} 
$$
$$ 
+ 2 \left( \nabla^\mu \nabla^\nu f(\phi)\right)R
      - 2 g^{\mu\nu} \left( \nabla^2f(\phi)\right)R
   - 4 \left( \nabla_\rho \nabla^\mu f(\phi)\right)R^{\nu\rho}
      - 4 \left( \nabla_\rho \nabla^\nu f(\phi)\right)R^{\mu\rho} 
$$
\be\lb{gb4b}
+ 4 \left( \nabla^2 f(\phi) \right)R^{\mu\nu}
+ 4g^{\mu\nu} \left( \nabla_{\rho} \nabla_\sigma f(\phi) \right) R^{\rho\sigma}
- 4 \left(\nabla_\rho \nabla_\sigma f(\phi) \right) R^{\mu\rho\nu\sigma}.
\ee
The equations (\ref{g3}) and (\ref{gb4b}) are the full system of equations describing the theory.

The following discussion is focused on the isotropic and homogeneous vacuums of the model 
with zero spatial curvature. The corresponding  distance element for these vacuums is given by
$$
g_4=-dt^2+a^2(t)\sum_{i=1}^3 dx_i^2.
$$
The formulas for the  Levi-Civita connection and the curvature of this background are well known, they are explicitly
$$
\Gamma^t_{ij}=a^2 H \delta_{ij}\ ,\qquad
\Gamma^i_{jt}=\Gamma^i_{tj}=H\delta^i_{\ j}\ ,
\qquad R_{itjt}=-\left(\dot H + H^2\right)\delta_{ij},
$$
$$
R_{ijkl}=a^4 H^2\left(\delta_{ik} \delta_{lj} - \delta_{il} \delta_{kj}\right),\qquad
R_{tt}=-3\left(\dot H + H^2\right)\ ,
$$
\be\lb{wk}
R_{ij}= a^2 \left(\dot H
+ 3H^2\right)\delta_{ij},\qquad R= 6\dot H + 12 H^2.
\ee
The other components are all zero. 

\section{Models without potential with flat spatial metric}

For the case $V(\phi)=0$, by assuming that the spatial curvature is $k=0$, the corresponding equations of motion (\ref{g3}) and (\ref{gb4b}) reduce to
\be\lb{1}
\frac{\dot{\phi}^2}{2}=3H^2(1+\dot{f}(\phi)H),
\ee
\be\lb{2}
\frac{\dot{\phi}^2}{2}=-2(H^2+\dot{H})(1+\dot{f}(\phi)H)-H^2(1+\ddot{f}(\phi)),
\ee
\be\lb{3}
\ddot{\phi}=-3H\dot{\phi}+3H^2 f'(\phi)(H^2+\dot{H}).
\ee
Our aim is to characterize the behavior of the solutions of these equations, without finding explicit solutions.

\subsection{General analysis}
The following analysis is focused on the two equations (\ref{1})-(\ref{2}). From equation (\ref{1}) it is immediately deduced that
\be\lb{polo}
\frac{\dot{\phi}^2}{6H^2}=1+\dot{f}H\geq0.
\ee
This implies that the inequality
$$
H f'(\phi)\dot{\phi}\geq-1,
$$
is satisfied during the whole evolution of the universe. On the other hand, the equation (\ref{2}) can be expressed by use of (\ref{1}) as follows
$$
\frac{\dot{\phi}^2}{2}=-2(H^2+\dot{H})(1+\dot{f}(\phi)H)-H^2(1+\ddot{f}(\phi))
$$
$$
=-2H^2(1+\dot{f}(\phi)H)-2\dot{H}(1+\dot{f}(\phi)H)-H^2(1+\ddot{f}(\phi))
$$
$$
=-\frac{\dot{\phi}^2}{3}-2\dot{H}(1+\dot{f}(\phi)H)-H^2(1+\ddot{f}(\phi))
$$
where in the last step (\ref{1}) was taken into account. The last identity is equivalent to 
$$
\frac{5\dot{\phi}^2}{6}+H^2=-2\dot{H}(1+\dot{f}(\phi)H)-H^2\ddot{f}(\phi).
$$
It may be easily seen that the right term is a total derivative, thus the last equality can be expressed as
$$
\frac{5\dot{\phi}^2}{6}+H^2=-2\frac{dH}{dt}-\frac{d(\dot{f}H^2)}{dt}\geq 0.
$$
In other words
$$
\frac{d}{dt}(2H+\dot{f}H^2)\leq 0.
$$
By integrating the last inequality the following bound is obtained
\be\lb{ready}
H(2+\dot{f}H)\leq C_0,\qquad t\geq0,
\ee
with $C_0$ being the  value of the quantity $H(2+\dot{f}H)$ at $t=0$. This bound is valid for times $t>0$. Furthermore, it can be easily generalized for two arbitrary times $t_1$ and $t_2$ such that $t_2>t_1$, with the left hand of the inequality referred to the time $t_2$ and the right hand to $t_1$. 

The inequality (\ref{ready}) derived above has important consequences. 
Assume for a moment that the initial condition is such that $C_0<0$. The inequality (\ref{polo}) shows that the factor $2+\dot{f}H\geq 1$, therefore
this case corresponds to $H< 0$. Thus, for this initial condition, the universe is contracting at $t=0$. By taking into account (\ref{ready}) and that $2+\dot{f}H\geq 1$ it follows that
\be\lb{ready2}
H\leq \frac{C_0}{2+\dot{f}H}\leq 0,\qquad t\geq0.
\ee
Thus the universe is always contracting in the future if it is contracting at $t=0$. This is an universal conclusion, no matter the form of the coupling 
$f(\phi)$.
Furthermore, for the past, the equation (\ref{ready}) is converted into
\be\lb{readyatras}
H\geq \frac{C_0}{2+\dot{f}H}\geq C_0,\qquad t\leq0,
\ee
Again, in the last step the inequality $2+\dot{f}H\geq 1$ was taken into account.

Suppose now that  the initial condition is $C_0>0$. Then
\be\lb{ready2}
H\leq \frac{C_0}{2+\dot{f}H}\leq C_0,\qquad t\geq0.
\ee
For the past, the equation (\ref{ready}) is converted into
\be\lb{readyatras}
H\geq \frac{C_0}{2+\dot{f}H}\geq 0,\qquad t\leq0,
\ee
Thus, if the universe is expanding at $t=0$, it was expanding always in the past. Again, this is an universal conclusion, without reference to the particular
form of the coupling $f(\phi)$. These results show in particular that there are no cyclic cosmologies for these models.

We have made the analogous analysis when a cosmological constant $\Lambda\neq 0$ is turned on. But we have obtained no universal conclusions in this case. 
In fact, when the curvature is turned on, the resulting inequalities are analogous to the ones above but with the replacement $C_0\to C_0+\Lambda t$.
In particular, cyclic cosmologies are allowed when the cosmological constant is turned on.

\subsection{The behavior of the scalar field}
Consider now the behavior of $\phi$ and $\dot{\phi}$. First,
by taking into account that $\dot{f}(\phi)=f'(\phi)\dot{\phi}$,  the equation (\ref{1}) becomes a quadratic algebraic relation for $\dot{\phi}$.
Its solution is 
\be\lb{c1}
\dot{\phi}=\frac{H}{2}\bigg[6H^2f'(\phi)\pm \sqrt{36H^4f'(\phi)^2+24}\bigg].
\ee
If the negative branch is chosen in (\ref{c1}) then
the time derivative is given by
\be\lb{c1d}
2\dot{\phi}=H\bigg[6H^2f'(\phi)-\sqrt{36H^4f'(\phi)^2+24}\bigg].
\ee
It is convenient to express this formula as
\be\lb{c5d}
2\dot{\phi}=6H^3f'(\phi)\bigg[1-\sqrt{1+\frac{24}{36H^4f'(\phi)^2}}\bigg].
\ee
By taking into account that $\sqrt{1+x}\sim 1+x/2$ for $x<<1$ and $\sqrt{1+x}\sim\sqrt{x}$ for $x>>1$, one can draw several conclusions. First, it follows from (\ref{c5d}) that if $f'(\phi)\to\pm\infty$ and $H$ is finite, then $\dot{\phi}\to 0$. Also, if $f'(\phi)\neq 0$ then $\dot{\phi}\to 0$ when $H\to\pm\infty$.
Instead, if $f'(\phi)=0$ then it may happen that $\dot{\phi}\to\pm\infty$ when $H\to \pm \infty$, as follows from (\ref{c1d}). For this reason, a coupling $f'(\phi)$ 
which never reaches a zero will  be chosen. An example of this may be a coupling $f'(\phi)>0$ with a global minimum $f'_m>0$. In addition, this condition implies that
$\dot{\phi}\to 0$ when $H\to 0$ and $f'(\phi)$ is finite, this is directly seen by use of (\ref{c5d}). 

Note however, that there may be an indetermination when $H\to 0$ and $f'(\phi)\to\pm \infty$. Consider first the case $H^2 f'(\phi)\to c$, with $c$ a constant. Then $H^3f'(\phi)\to 0$ and from (\ref{c5d})
it is clear that $\dot{\phi}\to 0$. If instead $H^2 f'(\phi)\to 0$ then $H^3f'(\phi)\to 0$ and again it is seen from (\ref{c5d}) and that  $\sqrt{1+x}\sim\sqrt{x}$ for $x>>1$ that $\dot{\phi}\to 0$ in this case.
Finally, when $H^2 f'(\phi)\to \pm \infty$ it is seen from the same formula and that  $\sqrt{1+x}\sim 1+x/2$ for $x<<1$ that $\dot{\phi}\to 0$ again. In other words, there is no way in which $\dot{\phi}\to\pm\infty$ if the coupling $f'(\phi)$ is never zero.

Now, if $\dot{\phi}$ is interpreted as a function of the two variables $(f'(\phi), H)$ given by (\ref{c1d}), then the properties shown above show that it vanishes
at the point $(f'_m,0)$ and at any point of the infinite. The fact that $\dot{\phi}$ is a continuous function shows that it must have a minimum and a maximum somewhere. In fact, the restriction
of this function to any straight line in the space $(f'(\phi), H)$ connecting the point $(f'_m,0)$ with some point of at the infinite interpolates between the two zeros continuously. The Bolzano theorem implies  the presence of a minimum or a maximum in any of these directions. These directions are parametrized
an "angular" coordinate $\vartheta$ and since the function $\dot{\phi}$ is well behaved,  these extrema varies continuously with the angle. As the angular coordinate $0\leq\vartheta<2\pi$ is compact,  it follows that there should exist a global minimum $\dot{\phi}_1$ and a global maximum $\dot{\phi}_2$. Therefore
$$
\dot{\phi}_2\leq\dot{\phi}\leq\dot{\phi}_1,
$$
and, by simple integration of the last expression, it follows that
$$
 \phi_0+\dot{\phi}_1t\leq\phi\leq\phi_0+\dot{\phi}_2t,\qquad t\geq 0,
$$
\be\lb{bondo}
 \phi_0+\dot{\phi}_2 t\leq\phi\leq\phi_0+\dot{\phi}_1t,\qquad t\leq 0.
\ee
This means that the values of $\phi$ are bounded by two linear time functions, and therefore are bounded for any finite time value $t$.

\section{The possibility of eternal universes}

Consider again the equations (\ref{1})-(\ref{3}) with the branch in which $\dot{\phi}$ is bounded, that is, the branch described by (\ref{c5d}). In the following, it will be assumed that $f'(\phi)$ and $f''(\phi)$ are not divergent for any finite value of $\phi$. Suppose that the universe falls into a singularity at a given finite time, which can be chosen when $t\to \pm 0$, by a shift of time. The choice $t\to 0$ is for simplicity, the singularity may be at any value $t_0$ by a choice of a convenient parametrization. 
Our aim is to find situations in which this assumption gives a contradiction. In these situations, the universe will be eternal.

The analysis given in the present section avoid some dubious arguments presented in \cite{yogo}, although similar conclusions are obtained. 
The dubious argument is the one below the formula (3.21) of that paper. We suspect that this formula may be a trivial one due to a numerical computer error. Thus, the present analysis will avoid such types of arguments. Since the results of this section are crucial, the calculations will be as  explicit as possible and will not rely in any computer algorithm. All the tools to be used in the following sections are all analytical and its validity can be seen directly. 

Assume first that the singularity in the curvature comes from the behavior $H\to\pm\infty$ when $t\to 0$, either from negative or positive times. Since one is working in the branch for which $\dot{\phi}$ is bounded, the first equation (\ref{1}) shows that
\be\lb{ci1}
\lim_{t\to 0}(1+H\dot{f})=0,
\ee
otherwise  $\dot{\phi}\to\pm \infty$, which is not in the selected branch. The last is a necessary condition, and implies that $\dot{\phi}\to 0$ at $t\to 0$
in such a way that $H\dot{\phi}\to -1/f'(\phi_0)$, $\phi_0$ being the value of $\phi$ at the singularity.

The third equation (\ref{3}), together with the fact that $H\dot{\phi}\to -1/f'(\phi_0)$ at $t\to 0$, imply that
\be\lb{ci3}
\lim_{t\to0}\ddot{\phi}= -\frac{3}{f'(\phi_0)}+\lim_{t\to 0}3H^2 f'(\phi)(H^2+\dot{H}).
\ee
In these terms, there are several situations to analyse. In all the following sections, the case $f'(\phi)>0$ or $f'(\phi)<0$ will be considered. The reason is that, otherwise, the bounds (\ref{bondo}) may not be satisfied, and this bounds are crucial for the following.
\\

\emph{The case of unbounded scalar field acceleration $\ddot{\phi}$ and $H\to\pm \infty$}
\\

A first possibility is that the acceleration $\ddot{\phi}$ is divergent at the singularity.
From (\ref{ci3}) it follows that this possibility is equivalent to 
\be\lb{2ci4}
\lim_{t\to 0}3H^2 f'(\phi)(H^2+\dot{H})\to \pm\infty.
\ee
From (\ref{2}), and by taking (\ref{1}) into account, it follows that $\dot{\phi}\to 0$ implies that
\be\lb{bondo}
2\dot{H}(1+H \dot{f})+H^2(1+\ddot{f})\to 0.
\ee
As it will be shown below, this is impossible to satisfy. As $\ddot{\phi}$ is divergent, so is $\ddot{f}=f'(\phi)\ddot{\phi}+f''(\phi)\dot{\phi}^2$, since it has been assumed that $f'(\phi)$ is never zero. 
But the term multiplying $\dot{H}$ in (\ref{bondo}) tends to zero. Thus, $\dot{H}$ should be much more divergent
that $H^2$, as it should cancel the divergent second term in (\ref{bondo}). This second term
diverges as $\ddot{\phi}H^2$, which clearly explode faster than $H^2$ if $\ddot{\phi}\to\pm \infty$. Thus, it follows that $\dot{H}>>H^2$ near the singularity and this, together with (\ref{3}) imply that
\be\lb{bondo2}
\ddot{\phi}\sim 3H^2f'(\phi_0)\dot{H}\to\pm \infty.
\ee
The last two relations (\ref{bondo})-(\ref{bondo2}) combine to give
\be\lb{hd}
2\dot{H}(1+H \dot{f})+3H^4f'(\phi_0)^2\dot{H}\to 0.
\ee
This limit is not true, since $1+H \dot{f}+3H^4f'(\phi_0)^2\to 3H^4f'(\phi_0)^2$ by (\ref{bindo}), and this quantity is divergent as $H^4$. Furthermore, this quantity is also multiplying $\dot{H}$ in (\ref{hd}), which is also highly divergent.
This contradiction shows that, for this branch, it is impossible to have a singularity with unbounded acceleration for $H\to\pm \infty$.  Thus, the only possibility is that
$H$ is finite.
\\

\emph{The case of unbounded scalar field acceleration $\ddot{\phi}$, $H^2<\infty$ and $\dot{H}\to\pm\infty$}
\\

Assume that $H\to H_0$ and that $\dot{H}$ is divergent. Then from (\ref{1})
it follows, for $t\to 0$, that
\be\lb{lidl}
\ddot{\phi}\sim 3H_0^2 f'(\phi_0)\dot{H}\to \pm\infty.
\ee
On the other hand (\ref{2}) can be expressed, by use of (\ref{1}) as follows
$$
\frac{\dot{\phi}^2}{2}=-(H^2+\dot{H})\frac{\dot{\phi}^2}{3H^2}-H^2(1+\ddot{f}(\phi)),
$$
This is equivalent to 
$$
\frac{5\dot{\phi}^2}{6}+H^2=-\frac{\dot{\phi}^2\dot{H}}{3H^2}-H^2\ddot{f}(\phi) 
$$
Thus, near the singularity
\be\lb{and}
 \frac{5\dot{\phi}_0^2}{6}+H_0^2=-\frac{\dot{\phi}_0^2\dot{H}}{3H_0^2}-3H_0^4 f«^2(\phi_0)\dot{H}-f''(\phi_0)H_0^2\dot{\phi}_0^2,
\ee
where in the last step the equality $\ddot{f}=f'(\phi)\ddot{\phi}+f''(\phi)\dot{\phi}^2$ was used, together with 
the asymptotic behavior (\ref{lidl}). Here $\dot{\phi}_0$ is the velocity of the scalar field, which is not necessarily zero in this case, but we are working in the branch in which it is finite. 
Now, the left hand of (\ref{and}) is bounded by definition, since $\dot{\phi}_0$ and $H_0$ are finite by our assumptions. But the right hand of (\ref{and}) can be expressed as
\be\lb{and2}
 \frac{5\dot{\phi}_0^2}{6}+H_0^2=-\bigg(\frac{\dot{\phi}_0^2}{3H_0^2}+3H_0^4 f^{'2}(\phi_0)\bigg)\dot{H}-f''(\phi_0)H_0^2\dot{\phi}^2.
\ee
 Both terms multiplying $\dot{H}$ in the right hand are positive defined, and it is impossible to fix a value of $H_0$ and $\phi_0$ for which they cancel. Thus, the right hand is divergent, but the left is not which is absurd. Thus, there does not exist such singularity in this type of models, in the chosen branch, if $H_0$ is different from zero. 

However, there is a further possibility, that is, that $H\to 0$ and $\dot{H}\to\pm \infty$. This limit is better
studied by looking directly to equations (\ref{1})-(\ref{3}).
 The equation (\ref{1}) for $H\to 0$ shows that $\dot{\phi}\to 0$. Thus $1+\dot{f}H\to 1$. The equation (\ref{2}) reduces to 
\be\lb{2ss}
-2\dot{H}-H^2\ddot{f}(\phi)\to 0,
\ee
and the equation (\ref{3})
\be\lb{3ss}
\ddot{\phi}\to 3H^2 f'(\phi)\dot{H}.
\ee
Clearly, there is an indetermination of the type $0.\infty$ in both equations. But taking into account that
$\ddot{f}=f'(\phi)\ddot{\phi}+f''(\phi)\dot{\phi}^2$ in (\ref{2ss}) and by taking into account (\ref{3ss}) and that $f''(\phi)$ is finite in this branch, it is obtained that 
\be\lb{2s3}
-2\dot{H}-H^2\ddot{f}(\phi)\to -2\dot{H}-3H^4 f'(\phi)^2\dot{H}\to -2\dot{H}\to \infty.
\ee
The contradiction between (\ref{2s3}) and (\ref{2ss}) shows that this regime also does not exist.
\\

\emph{The case with bounded scalar field acceleration $\ddot{\phi}$}
\\

A further possibility is that the limit (\ref{ci3}) is finite, which means that the acceleration $\ddot{\phi}$ is bounded. This implies that
\be\lb{ci4}
\lim_{t\to 0}3H^2 f'(\phi)(H^2+\dot{H})=l_0,
\ee
where $l_0$ is a finite number. There are three possibilities to consider. One is that $H\to 0$ and that $\dot{H}\to\pm\infty$ in such a way that (\ref{ci4}) holds. But this possibility is easily ruled out as follows.
From (\ref{1}) it is seen that $\dot{\phi}\to 0$. Thus, the equation (\ref{2}) and the assumption of bounded acceleration imply that $\dot{H}\to 0$, which is a contradiction. The second possibility is that $H$ is finite and $\dot{H}\to\pm\infty$. But this clearly does not satisfy (\ref{ci4}). In addition, if $H$ and $\dot{H}$ are finite, there is no singularity and the universe is eternal. Thus, the only possibility for having a singular curvature is that $H\to\pm \infty$ and $\dot{H}\to \pm \infty$. By taking into account this, the last equation gives the following necessary condition 
\be\lb{ci5}
\lim_{t\to 0}(H^2+\dot{H})=0.
\ee
On the other hand, from (\ref{1}) and the conclusion that $H\to\pm \infty$, it is seen that for this case $H\dot{\phi}\to -1/f'(\phi_0)$, thus in particular $\dot{\phi}\to 0$ at the singular point $t\to 0$. This, combined with (\ref{2}) and (\ref{ci5}) leads to the conclusion that
\be\lb{ci6}
\lim_{t\to  0}H^2(1+\ddot{f}(\phi))=0.
\ee
Thus, another necessary condition is that
\be\lb{ci60}
\lim_{t\to  0}\ddot{f}(\phi)=-1.
\ee
Since $\ddot{f}=f'(\phi)\ddot{\phi}+f''(\phi)\dot{\phi}^2$ and $\dot{\phi}\to 0$ at $t\to 0$, the last condition is equivalent to
\be\lb{ci61}
\lim_{t\to  0}\ddot{\phi}=-\frac{1}{f'(\phi_0)}.
\ee
This, combined with the formulas (\ref{ci3}) and (\ref{ci4}) shows that 
\be\lb{ci7}
\lim_{t\to 0}H^2 (H^2+\dot{H})=\frac{2}{3f'^2(\phi_0)}.
\ee
The last condition puts several restrictions on the behavior
 near the singularity. First, note that the solution of $\dot{H}+H^2=0$ is $H=c/t$. This suggest that it is convenient
to parametrize $H$ near the singularity as
\be\lb{ci8}
H=\frac{c}{t+h(t)},
\ee
with $h(t)$ being a function of time that goes to zero faster than linearly. The condition (\ref{ci7}) is then
\be\lb{ci9}
\lim_{t\to 0}\frac{c^2-c-ch'(t)}{(t+h(t))^4}=\frac{2}{3c^2 f'^2(\phi_0)}.
\ee
As $h(t)$ goes to zero faster than linearly, the last equation can be satisfied only if $c=1$ and 
\be\lb{ci10}
h'(t)=-\frac{2}{3f'^2(\phi_0)}t^4(1+g(t)).
\ee
with $g(t)$ a function that goes to zero at $t\to 0$, not necessarily analytical. Therefore
\be\lb{ci11}
h(t)=-\frac{2}{15f'^2(\phi_0)}t^5+m(t).
\ee
Here $m(t)$ goes to zero faster than $t^5$, and is not necessarily analytic. The conclusion is that, near the singularity
\be\lb{ci12}
H=\frac{1}{t-\frac{2}{15f'^2(\phi_0)}t^5+m(t)}.
\ee
Another way to justify this expression is to postulate that 
$$
H=\frac{1}{q(t)},
$$
with $q(t)$ going to zero at $t\to 0$.
The condition $H^2(\dot{H}+H^2)\to cte$ at $t\to 0$ becomes
\be\lb{cil}
\lim_{t\to 0}\frac{1-q'(t)}{q(t)^4}=\frac{2}{3f'^2(\phi_0)}.
\ee
Thus $q'(t)=1+l(t)$ with $l(t)\to 0$ when $t\to 0$.
Therefore $q(t)=t+k(t)$ with $k(t)\to 0$ when $t\to 0$ faster than linearly.
However (\ref{cil}) shows that $k'(t)/t^4$ tends to a constant. Integration of this expression will lead again to (\ref{ci12}).

Now, the expression (\ref{ci12}) combined with (\ref{ci6}) and (\ref{ci60}) imply that
\be\lb{ci13}
\ddot{f}(\phi)=-1+\alpha t^{2+\epsilon}+s(t),
\ee
with $\epsilon>0$ and $s(t)$ containing the terms that go to zero even faster than $t^{2+\epsilon}$. One may consider the possibility
that $\ddot{f}$ goes to zero not like any power law, for instance as $\ddot{f}(\phi)=-1+\alpha t^{2}u(t)$ with $u(t)\to 0$ as $t\to 0$ and non analytical. But we will argue below that this is not the case. 
The last condition may be integrated to give
\be\lb{ci15}
\dot{f}(\phi)=-t+\frac{\alpha t^{3+\epsilon}}{3+\epsilon} +r(t),
\ee
where $r(t)$ goes to zero faster than $t^{3+\epsilon}$. Finally, the fact that
\be\lb{ci14}
\ddot{\phi}=-\frac{1}{f'(\phi_0)},\qquad \to \qquad \dot{\phi}=-\frac{t}{f'(\phi_0)}+w(t),
\ee
where $w(t)$ goes to zero faster than linearly. All the obtained expressions are valid in an small interval near $t=0$.

There are further consequences that can be drawn from the obtained expressions. By taking into account (\ref{ci12}), (\ref{ci13}) and (\ref{ci14}) the equation (\ref{1}), namely 
$$
\frac{\dot{\phi}^2}{2}=3H^2(1+\dot{f}(\phi)H),
$$
is converted near the singularity into
$$
\frac{1}{6}\bigg(t-\frac{2}{15f'^2(\phi_0)}t^5+m(t)\bigg)^3\bigg(\frac{t}{f'(\phi_0)}-w(t)\bigg)^2=-\frac{2}{15f'^2(\phi_0)}t^5+\frac{\alpha t^{3+\epsilon}}{3+\epsilon} +m(t)+r(t).
$$
But the coefficients proportional to $t^5$ match only if 
$$
\alpha=\frac{3}{2f'(\phi_0)^2},\qquad \epsilon=2.
$$
This matching is in fact the justification for proposing that 
$\ddot{f}(\phi)=-1+\alpha t^{2+\epsilon}+s(t)$ near the singularity
instead of a generic $\ddot{f}(\phi)=-1+\alpha t^{2}u(t)$ with $u(t)\to 0$ as $t\to 0$.
Without this dependence, there will not be matching between the quintic terms.
Now, with the value of $\alpha$ just found it follows by integrating 
$\ddot{f}(\phi)=-1+\alpha t^{2+\epsilon}+s(t)$ with respect to time that
\be\lb{ci16}
\dot{f}(\phi)=-t+\frac{3t^{5}}{10 f'(\phi_0)^2} +r(t),
\ee
where $r(t)$ goes faster than $t^5$ and is not necessarily analytic.

It should be emphasized that the matching of the quintic terms are necessary, but they do not insure the existence of a solution.
In fact, it is possible already to derive some inconsistencies which suggest the existence of a huge class of models
for which there are eternal solutions.
The last condition may be combined with (\ref{ci14}) to get further consequences which are impossible to satisfy. The approximation (\ref{ci14})
may be integrated to give
\be\lb{ci17}
\phi=\phi_0+\delta \phi(t)=\phi_0-\frac{t^2}{2f'(\phi_0)}+W(t),
\ee
where $W(t)$ is the primitive of $w(t)$, and grows faster than quadratically. Both (\ref{ci16}) and (\ref{ci17}) combined give that
$$
\dot{f}(\phi(t))\simeq f'(\phi(t))\dot{\phi}(t)\simeq\bigg[f'(\phi_0)+f''(\phi_0)\delta \phi(t)\bigg]\dot{\phi}(t)\simeq -t+\frac{3t^{5}}{10 f'(\phi_0)^2},
$$
or explicitly
$$
\bigg[f'(\phi_0)+f''(\phi_0)\bigg(-\frac{t^2}{2f'(\phi_0)}+W(t)\bigg)\bigg]\bigg(-\frac{t}{f'(\phi_0)}+w(t)\bigg)\simeq -t+\frac{3t^{5}}{10 f'(\phi_0)^2},
$$
up to higher order terms. Now, the left hand side may be expanded to obtain
$$
-t+f'(\phi_0) w(t)+\frac{f''(\phi_0)t^3}{2f'(\phi_0)^2}-\frac{f''(\phi_0)t^2w(t)}{2f'(\phi_0)}-\frac{f''(\phi_0)tW(t)}{f'(\phi_0)}
$$
\be\lb{flash}
+f(\phi_0)w(t)W(t)\simeq -t+\frac{3t^{5}}{10 f'(\phi_0)^2}. 
\ee
The linear terms clearly match. But there are problems to match the other terms. In order to see this, one should take into account that $w(t)$ grows faster than linearly, and that its primitive $W(t)$ goes faster than quadratically. Now, the third term in (\ref{flash}) is cubic, and since the right hand does not have any cubic term, it should be cancelled somehow. It can not be cancelled by the fourth or the fifth term, since the behavior
 of $w(t)$ or $W(t)$ described above makes these terms of higher order than three. The sixth term also is of higher order. But it can be cancelled by second term by assuming that
$$
w(t)=-\frac{f''(\phi_0)t^3}{2f'(\phi_0)^3}+w_2(t),\qquad \longrightarrow \qquad W(t)=-\frac{f''(\phi_0)t^4}{8f'(\phi_0)^3}+W_2(t).
$$
On the other hand, $w_2(t)$ can not include a quadratic term. If that were the case then the second term in (\ref{flash}) would contain a quadratic term that can not be compensated, as $V(t)$ contains at least cubic terms. Thus, $w_2(t)$ goes to zero faster than $t^3$ and therefore $W_2(t)$ goes to zero faster than $t^4$. From here it is seen that the fourth and the fifth term of (\ref{flash}) go like
$$
-\frac{f''(\phi_0)t^2v(t)}{2f'(\phi_0)}-\frac{f''(\phi_0)tV(t)}{f'(\phi_0)}\simeq \frac{f''(\phi_0)^2}{4f'(\phi_0)^4}t^5+\frac{f''(\phi_0)^2}{8f'(\phi_0)^4}t^5,
$$
up to higher order terms. This term match the quintic term of (\ref{flash}) if and only if
$$
\frac{3f''(\phi_0)^2}{8f'(\phi_0)^4}=\frac{3}{10 f'(\phi_0)^2}.
$$
This gives the following numerical relation defining $\phi_0$
\be\lb{condicion}
 f''(\phi_0)^2=\frac{4f'(\phi_0)^2}{5}.
 \ee
This is one of the mandatory conditions to be satisfied. However,
there exist a lot of functions for which (\ref{condicion}) is never satisfied. For instance, there are a lot
of couplings for which
$$
 0 \leq f''(\phi)\leq \frac{2f'(\phi)}{\sqrt{5}}.
$$
This is satisfied for instance for 
$$
0<f'(\phi)\leq c \exp(\frac{2}{\sqrt{5}}\phi).
$$
There are plenty of models that satisfy this constraint, as a bound by an exponential is not a very restrictive condition. Thus, for any of these models, for the coupling is bounded by 
an exponential, the cosmological solutions corresponding to our branch will be eternal.

\section{Models with flat spatial metric and cosmological constant  $\Lambda>0$ turned on}

For the case with a cosmological constant $\Lambda>0$ turned on, by assuming that the spatial curvature is $k=0$, the corresponding equations of motion are given by
\be\lb{1c}
\frac{\dot{\phi}^2}{2}+\Lambda=3H^2(1+\dot{f}(\phi)H),
\ee
\be\lb{2c}
\frac{\dot{\phi}^2}{2}-\Lambda=-2(H^2+\dot{H})(1+\dot{f}(\phi)H)-H^2(1+\ddot{f}(\phi)),
\ee
\be\lb{3c}
\ddot{\phi}=-3H\dot{\phi}+3H^2 f'(\phi)(H^2+\dot{H}).
\ee
In analogous way than in equations (\ref{c1})-(\ref{bondo}), it may be show
that the negative branch of the scalar field $\dot{\phi}$ is bounded in this case. Based on this, the following analysis can be done.
\\

\emph{The case of unbounded scalar field acceleration $\ddot{\phi}$ and $H\to\pm \infty$}
\\

As before, consider the  possibility is that the acceleration $\ddot{\phi}$ is divergent at the singularity.
This possibility is equivalent to 
\be\lb{romi}
\lim_{t\to 0}3H^2 f'(\phi)(H^2+\dot{H})\to \pm\infty.
\ee
Assume that $H\to\pm \infty$ at the singularity $t\to 0$, either from the left or the right. As $\dot{\phi}$ is bounded then (\ref{1c}) holds only if 
\be\lb{bindo}
1+H\dot{f}\to 0,\qquad \dot{\phi}H\to -\frac{1}{f'(\phi_0)},
\ee
with $\phi_0$ is the value of $\phi$ at the singularity. 
Since $H^2\to \infty$ it is clear that $\dot{\phi}\to 0$. From (\ref{2c}), and by taking (\ref{1c}) into account, it follows that $\dot{\phi}$ is bounded only if
\be\lb{bondolar}
2\dot{H}(1+H \dot{f})+H^2(1+\ddot{f})\to \frac{\Lambda}{3}.
\ee
By use of $\ddot{f}=f'(\phi)\ddot{\phi}+f''(\phi)\dot{\phi}^2$, and by use of analogous arguments than the ones below (\ref{bondo}), it is obtained that $\dot{H}>>H^2$ near the singularity and this, together with (\ref{3c}) permits to conclude that
 \be\lb{bondol2}
\ddot{\phi}\sim 3H^2f'(\phi_0)\dot{H}\to\pm \infty,
\ee
at the singularity. The last two relations (\ref{bindo})-(\ref{bondolar}) combine to give
\be\lb{hdll}
2\dot{H}(1+H \dot{f})+3H^4f'(\phi_0)^2\dot{H}\to \frac{\Lambda}{3}.
\ee
However, this limit can not be true, since $1+H \dot{f}+3H^4f'(\phi_0)^2\to 3H^4f'(\phi_0)^2$ by (\ref{bindo}), and this quantity is divergent as $H^4$. Furthermore, this quantity is also multiplying $\dot{H}$ in (\ref{hd}), which is also highly divergent.
This contradiction shows that, for this branch, it is impossible to have a singularity with unbounded acceleration, unless $H$
is finite. 
\\

\emph{The case of unbounded scalar field acceleration $\ddot{\phi}$, $H^2<\infty$ and $\dot{H}\to\pm\infty$}
\\

Consider now the possibility that $H\to H_0$ and that $\dot{H}$ is divergent. Then from (\ref{1c})
it follows, for $t\to 0$, that
\be\lb{romi2}
\ddot{\phi}\sim 3H_0^2 f'(\phi_0)\dot{H}\to \pm\infty.
\ee
On the other hand (\ref{2c}) can be expressed, by use of (\ref{1c}) as follows
$$
\frac{\dot{\phi}^2}{2}-\Lambda=-\frac{(H^2+\dot{H})(\dot{\phi}^2+2\Lambda)}{3H^2}-H^2(1+\ddot{f}(\phi)),
$$
Therefore
$$
\frac{5\dot{\phi}^2}{6}+H^2-\frac{\Lambda}{3}=-\frac{(\dot{\phi}^2+2\Lambda)\dot{H}}{3H^2}-H^2\ddot{f}(\phi) 
$$
Thus, near the singularity
\be\lb{and}
 \frac{5\dot{\phi}_0^2}{6}+H_0^2-\frac{\Lambda}{3}=-\frac{(\dot{\phi}_0^2+2\Lambda)\dot{H}}{3H_0^2}-3H_0^4 f«^2(\phi_0)\dot{H}-f''(\phi_0)H_0^2\dot{\phi}^2,
\ee
where in the last step the equality $\ddot{f}=f'(\phi)\ddot{\phi}+f''(\phi)\phi^2$ was used, together with 
the asymptotic behavior (\ref{romi2}). Here $\dot{\phi}_0$ is the velocity of the scalar field, which is not necessarily zero in this case, but we are working in the branch in which it is finite. 
Now, the left hand of (\ref{and}) is bounded by definition, since $\dot{\phi}$ and $H_0$ are finite by our assumptions. But the right hand of (\ref{and}) can be expressed as
\be\lb{and2}
 \frac{5\dot{\phi}_0^2}{6}+H_0^2-\frac{\Lambda}{3}=-\bigg(\frac{\dot{\phi}_{0}^{2}+2\Lambda}{3H_0^2}+3H_0^4 f'(\phi_0)^2\bigg)\dot{H}-f''(\phi_0)H_0^2\dot{\phi}^2.
\ee
 Both terms multiplying $\dot{H}$ in the right hand are positive defined, and it is impossible to fix a value of $H_0$ and $\phi_0$ for which they cancel. Thus, the right hand is divergent, but the left is not which is absurd. Thus, there does not exist such singularity in this type of models, in the chosen branch, if $H_0$ is different from zero. 

However, there is a further possibility, that is, that $H\to 0$ and $\dot{H}\to\pm \infty$. But it is easy to see from equations (\ref{1c})-(\ref{3c}) that the case $H=0$ is not allowed when the cosmological constant $\Lambda$ is turned on. Thus, for divergent acceleration $\ddot{\phi}\to\pm \infty$ there is no singularity in this branch.
\\

\emph{The case with bounded scalar field acceleration $\ddot{\phi}$}
\\

Next, consider the possibility that $\ddot{\phi}$ is finite. The equation (\ref{3c}) is the same as the one without cosmological constant,
since the derivative of $\Lambda$ is simply zero. In view of this, the discussions given in the previous sections show that $H^2+\dot{H}\to 0$ near the singularity, and furthermore both $H^2$ and $\dot{H}$ are divergent at the singular point. It is not difficult to check that the condition (\ref{ci4}) and (\ref{ci5}) that was obtained for the case $\Lambda=0$ also hold in this case. In addition (\ref{1c}) and the fact that $H^2\to\infty$ shows that $H\dot{\phi}\to-1/f'(\phi_0)$
and therefore $\dot{\phi}\to 0$ near the singularity. These conditions are similar to the ones found for $\Lambda=0$. The main difference follows from equation (\ref{2c}). In fact, by taking into account that  $H^2+\dot{H}\to 0$, it is seen that equation (\ref{2c}) becomes
\be\lb{cil6}
\lim_{t\to  0}H^2(1+\ddot{f}(\phi))=\Lambda.
\ee
This is different that condition (\ref{ci6}) and in fact, it reduces to that case only when $\Lambda\to 0$.
The resulting necessary condition is that
\be\lb{cil60}
\lim_{t\to  0}\ddot{f}(\phi)=-1.
\ee
As $\ddot{f}=f'(\phi)\ddot{\phi}+f''(\phi)\dot{\phi}^2$ and $\dot{\phi}\to 0$ at $t\to 0$, the last condition implies
\be\lb{cil61}
\lim_{t\to  0}\ddot{\phi}=-\frac{1}{f'(\phi_0)}.
\ee
This, combined with $H\dot{\phi}\to-1/f'(\phi_0)$ and the formulas (\ref{ci3}) and (\ref{ci4})\footnote{Which are also valid in this case, as the equation of motion for $\phi$ are unchanged by the presence of a cosmological constant. Note that in presence of a potential $V(\phi)$ there appears a term proportional to $V'(\phi)$, but for a constant potential (a cosmological constant) this term do not contribute.} shows that 
\be\lb{cil7}
\lim_{t\to 0}H^2 (H^2+\dot{H})=\frac{2}{3f'^2(\phi_0)}.
\ee
This is exactly the condition (\ref{ci7}) obtained in the previous section, where it was shown
to lead to (\ref{ci12}). Thus formula (\ref{ci12}) applies in this case. The formula (\ref{ci12}), combined
 with (\ref{cil6}) and (\ref{cil60}) imply that
\be\lb{cil13}
\ddot{f}(\phi)=-1+\Lambda t^2+\alpha t^{2+\epsilon}+s(t),
\ee
with $s(t)$ containing the terms that go to zero faster than quadratically. Thus
\be\lb{cil15}
\dot{f}(\phi)=-t+\frac{\Lambda}{3}t^3+\frac{\alpha t^{3+\epsilon}}{3+\epsilon} +r(t),
\ee
where $r(t)$ goes to zero faster than $t^{3}$. The formula (\ref{ci14}) is also unchanged.
By taking into account
$$
\frac{\dot{\phi}^2}{2}+\Lambda=3H^2(1+\dot{f}(\phi)H),
$$
it is seen that, near the singularity, the following relation holds
\be\lb{33}
\frac{1}{6}\bigg(t-\frac{2}{15f'^2(\phi_0)}t^5+m(t)\bigg)^3\bigg[\bigg(\frac{t}{f'(\phi_0)}-w(t)\bigg)^2+2\Lambda\bigg]=-\frac{2}{15f'^2(\phi_0)}t^5+\frac{\Lambda t^{3}}{3}+\frac{\alpha t^{3+\epsilon}}{3+\epsilon} +m(t)+r(t).
\ee
The coefficients proportional to $t^3$ do match. The quintic terms match when
$$
\alpha=\frac{3}{10 f'(\phi_0)^2},
$$
and this, together with (\ref{cil15}) imply that
\be\lb{ci1o6}
\dot{f}(\phi)=-t+\frac{\Lambda}{3}t^3+\frac{3t^{5}}{10 f'(\phi_0)^2} +r(t),
\ee
where $r(t)$ goes faster than $t^5$ and is not necessarily analytic. However, arguments analogous to the ones giving (\ref{flash})
show that 
$$
-t+f'(\phi_0) w(t)+\frac{f''(\phi_0)t^3}{2f'(\phi_0)^2}-\frac{f''(\phi_0)t^2w(t)}{2f'(\phi_0)}-\frac{f''(\phi_0)tW(t)}{f'(\phi_0)}
$$
\be\lb{flash2}
+f(\phi_0)w(t)W(t)\simeq -t+\frac{\Lambda}{3}t^3+\frac{3t^{5}}{10 f'(\phi_0)^2}, 
\ee
should be satisfied. This reduces to (\ref{flash}) when $\Lambda\to 0$. The terms in (\ref{flash2}) match by postulating that
$$
w(t)=-\alpha t^3+w_2(t),\qquad W(t)=-\frac{\alpha t^4}{4}+W_2(t)
$$
which gives that two linear conditions whose solution is
\be\lb{cierra}
\alpha=-\frac{2}{5 f'(\phi_0)f^{''}(\phi_0)},\qquad -\frac{2}{5 f'(\phi_0)f^{''}(\phi_0)}+\frac{f^{''}(\phi_0)}{2f'^{2}(\phi_0)}=\frac{\Lambda}{3}.
\ee
Note that the second condition (\ref{cierra}) coincides with (\ref{condicion}) when $\Lambda\to 0$, which 
is a good consistency test. Thus, the second (\ref{cierra})  can be interpreted as a generalization of (\ref{condicion}). There are a lot
of couplings which do not satisfy this second condition, as for the case $\Lambda=0$. For these models we have again eternal solutions, when the scalar field is in the negative branch.

\section{Spatial curvature $k=\pm 1$ turned on}
The equations in this case are given by
\be\lb{ecu1}
\frac{\dot{\phi}^2}{2}=3(H^2+\frac{k}{a^2})(1+\dot{f}(\phi)H),
\ee
\be\lb{ecu2}
\frac{\dot{\phi}^2}{2}=-2(H^2+\dot{H})(1+\dot{f}(\phi)H)-(H^2+\frac{k}{a^2})(1+\ddot{f}(\phi)),
\ee
\be\lb{ecu3}
\ddot{\phi}=-3H\dot{\phi}+3(H^2+\frac{k}{a^2}) f'(\phi)(H^2+\dot{H}).
\ee
The equation (\ref{ecu2}) can be worked out further by adding and subtracting a term proportional to $k/a^2$, the result is
\be\lb{wecu2}
-\frac{\dot{\phi}^2}{2}=2(1+\dot{f}H)(H^2+\frac{k}{a^2})+2(1+\dot{f}H)(\dot{H}-\frac{k}{a^2})+(1+\ddot{f})(H^2+\frac{k}{a^2}).
\ee
The first term of the right hand side is proportional to $\phi^2$, this follows directly from (\ref{ecu1}).
Thus, (\ref{wecu2}) becomes
\be\lb{wecu3}
-\frac{5\dot{\phi}^2}{6}=2(1+\dot{f}H)(\dot{H}-\frac{k}{a^2})+(1+\ddot{f})(H^2+\frac{k}{a^2}).
\ee
By taking into account the definition $H=\dot{a}/a$ it follows that some terms can be arranged as a total derivative as follows
\be\lb{wecu4}
-\frac{5\dot{\phi}^2}{6}-H^2=\frac{d}{dt}\bigg(2H+H^2\dot{f}+\frac{k\dot{f}}{a^2}\bigg)-\frac{k}{a^2}.
\ee
The important point is that the left hand side is obviously negative, thus the right hand is negative as well. Therefore
$$
\frac{d}{dt}\bigg(2H+\dot{f}(H^2+\frac{k}{a^2})\bigg)-\frac{k}{a^2}\leq 0.
$$
Integration of the last inequality gives
\be\lb{in1}
2H+\dot{f}(H^2+\frac{k}{a^2})-\int_0^t\frac{k}{a^2(\xi)}d\xi\leq C_0.
\ee
Here $C_0$ is the value of the quantity $2H+H^2\dot{f}+k\dot{f}/a^2$ at $t=0$. The left hand of the inequality is evaluated at a given future time, which is denoted by $t$.

The inequality (\ref{in1}) is an important constraint for the model. In order to visualize its consequences, assume that at $t$ the Hubble constant $H$
is positive $H>0$ and that it is approaching a singularity at $t_0>t>0$. Suppose that the singularity comes from a behavior
 of the lapse function
of the form $a(t)\sim c(t_0-t)^\alpha$.  Then $H\sim\alpha/(t-t_0)$, and clearly  $H>0$ only if $\alpha<0$, since $t<t_0$. It is convenient to express (\ref{in1}) as follows
\be\lb{win2}
2H+\dot{f}(H^2+\frac{k}{a^2})-\int_0^t\frac{k}{a^2(\xi)}d\xi\leq C_0.
\ee
By assumption $H>0$, thus by multiplying the whole inequality (\ref{win2}) by $H$ gives
 \be\lb{win3}
2H^2+H\dot{f}(H^2+\frac{k}{a^2})-H\int_0^t\frac{k}{a^2(\xi)}d\xi\leq C_0 H.
\ee
Now, the addition of the term $H^2+\frac{k}{a^2}$ to the last expression shows that
\be\lb{win4}
2H^2+(1+H\dot{f})(H^2+\frac{k}{a^2})-H\int_0^t\frac{k}{a^2(\xi)}d\xi\leq C_0 H+H^2+\frac{k}{a^2}.
\ee
By use of (\ref{ecu1}) it is directly seen that the second term of (\ref{win4}) is proportional
to $\dot{\phi}^2$. Thus it follows that (\ref{win4}) is given by
\be\lb{win5}
H^2-H\int_0^t\frac{k}{a^2(\xi)}d\xi\leq C_0 H+\frac{k}{a^2}-\frac{\dot{\phi}^2}{6}\leq C_0 H+\frac{k}{a^2}.
\ee
The inequality (\ref{win5}) reduces to 
\be\lb{win6}
H^2-\frac{k}{a^2}-H\int_0^t\frac{k}{a^2(\xi)}d\xi\leq C_0 H.
\ee
Our assumption is that  $a(t)\sim c(t_0-t)^\alpha$ with $\alpha<0$, thus the scale factor is divergent at this stage.
 It is implicitly assumed that $t_0$ is the first singularity after $t=0$. The inequality (\ref{win6}) then becomes
\be\lb{win7}
\frac{\alpha^2}{(t_0-t)^{2}}-kc^2(t_0-t)^{2\alpha}-\frac{\alpha}{t_0-t} I\leq C_0 \frac{\alpha}{t_0-t}.
\ee
Here $I$ is the value of the integral, which is finite since $1/a^2(t)$ does not have any singularity.
But the important point to remark is the following. The first term on the left hand is positive and has the singular behavior
 $1/(t_0-t)^2$, which is more explosive  than the term $1/(t_0-t)$  of the right hand. The only way that the inequality (\ref{win7}) can be fulfilled is that the second and the third term cancel this behavior. But clearly, none of them can do the job. Note that, during all the reasoning above, the value of $C_0$, the value of $k$ or the behavior
 of the coupling $f(\phi)$ was immaterial.

Now, consider the case $H<0$ in the past and falling into a singularity $H\sim \frac{\alpha}{t-t_0}$ at $t=t_0<0$ due to a behavior $a(t)\sim c(t-t_0)^{\alpha}$ with $\alpha<0$. For past times $t<0$ the inequality obtained by integrating (\ref{in1}) is given by
\be\lb{ine1}
C_0\leq \int_{t}^0\frac{2k}{a^2(\xi)}d\xi+ 2H+\dot{f}(H^2+\frac{k}{a^2}).
\ee
By multiplying by $H$ and remembering that $H<0$ it is found that
\be\lb{ine2}
H\int_{t}^0\frac{2k}{a^2(\xi)}d\xi+ 2H^2+H\dot{f}(H^2+\frac{k}{a^2})\leq C_0 H.
\ee
As before, the addition of the term $H^2+\frac{k}{a^2}$ converts this expression into 
\be\lb{ine3}
H\int_{t}^0\frac{2k}{a^2(\xi)}d\xi+ 2H^2\leq C_0 H+H^2+\frac{k}{a^2}-\frac{\dot{\phi}^2}{4}.
\ee
Thus, in particular,
\be\lb{ine4}
H\int_{t}^0\frac{2k}{a^2(\xi)}d\xi+ H^2-\frac{k}{a^2}\leq C_0 H.
\ee 
By choosing $k=1$ it is seen that the integral now is not divergent, since the function $a(t)$ do not have zeros.
Denote its value as $I$. Then the last bound is
\be\lb{ine6}
-\frac{k \alpha I}{t+t_0}+\frac{\alpha^2}{(t+t_0)^{2}}-\frac{k(t+t_0)^{2\alpha}}{c^2}\leq -\frac{C_0\alpha}{t_0+t}.
\ee 
All the terms proportional to $(t+t_0)^{2\alpha}$ vanish near the singularity. It is impossible to satisfy this inequality, since the
term $\frac{\alpha^2}{(t+t_0)^{2}}$ is the most explosive one and is positive, thus the left hand side of the inequality is larger than the right hand
side near a singularity, a clear contradiction. This is valid for $k=\pm 1$ and $k=0$. In short terms, by denoting $\beta=-\alpha$, the conclusions made about the past and future behavior can be 
stated as follows.
\\

\emph{Proposition:} The Gauss-Bonnet cosmology without potential and with spatial curvature $k=\pm 1$ or $k=0$ does not admit solutions for which there is a regime
falling into a singularity of the form $a(t)\sim c/(t_0-t)^\beta$, with $\beta>0$, neither in the past or future, no matter the explicit form of the coupling $f(\phi)$.
\\

We have also considered the other two complementary cases, namely $H<0$ falling into a power law in the past and 
$H<0$ falling into a singularity in the future. But the bounds that we found depend on the behavior of $\dot{\phi}$ and we can find no conclusions in this case.

\subsection{The negative branch of the scalar field}

As for the flat case, the equation (\ref{ecu1}) becomes a quadratic algebraic relation for $\dot{\phi}$.
Its solution is 
\be\lb{ce1}
\dot{\phi}=6H\bigg(H^2+\frac{k}{a^2}\bigg)f'(\phi)\pm \sqrt{36H^2\bigg(H^2+\frac{k}{a^2}\bigg)^2f'(\phi)^2+24\bigg(H^2+\frac{k}{a^2}\bigg)}.
\ee
If the negative branch is chosen in (\ref{ce1}) 
\be\lb{c5d}
\dot{\phi}=6H\bigg(H^2+\frac{k}{a^2}\bigg)f'(\phi)\bigg[1-\sqrt{1+\frac{24}{36H^2\bigg(H^2+\frac{k}{a^2}\bigg)f'(\phi)^2}}\bigg].
\ee
The analysis goes essentially as in the case with $k=0$, if the coupling $f'(\phi)$ is never zero, but there are more cases to take into account. First, if $H\to H_0$, $f'(\phi)\to f'(\phi_0)$
and $a\to 0$, then (\ref{c5d}) shows that $\dot{\phi}$ is finite. The same happens when $H\to H_0$, $f'(\phi)\to f'(\phi_0)$
and $a\to \infty$. When $a\to a_0$, $f'(\phi)\to f'(\phi_0)$ and $H\to 0$ one has that $\dot{\phi}^2<\infty$.
Also, when $a\to a_0$, $f'(\phi)\to f'(\phi_0)$ and $H\to \infty$ it is obtained that $\dot{\phi}\to 0$. In the case 
when $a\to a_0$, $f'(\phi)\to\pm \infty$ and $H\to H_0$ also  $\dot{\phi}\to 0$.

Similar conclusions are obtained when $H\to 0$, $a\to\infty$ and $f'(\phi)\to f'(\phi_0)$, or when $H\to \infty$, $a\to 0$ and $f'(\phi)\to f'(\phi_0)$ or even when $H\to \infty$, $a\to\infty$ and $f'(\phi)\to f'(\phi_0)$. 
From (\ref{c5d}) it is seen that if $f'(\phi)\to\pm\infty$ and 
$$
H^2\bigg(H^2+\frac{k}{a^2}\bigg)\to c,
$$
with $c$ a constant, then $\dot{\phi}\to 0$. Also, if $f'(\phi)\neq 0$ then $\dot{\phi}\to 0$ when $H\to\pm\infty$ and $a$ is finite and non zero. The same holds when $H\to\pm \infty$ and $a$ goes to zero or infinite. 
In addition, if $k=-1$ and $(H^2+\frac{k}{a^2})\to 0$, with $H$ and $a$ finite, then $\dot{\phi}\to 0$ as well. 
In addition when $H\to 0$  it may be possible to have 
$$
H\bigg(H^2+\frac{k}{a^2}\bigg)\to \pm \infty,\qquad H^2\bigg(H^2+\frac{k}{a^2}\bigg)\to c,
$$
with $c$ a constant. In this case, $\dot{\phi}\to \pm\infty$. But this limit implies that $H\to 0$ and $H^2/a^2\to c'$, with $c'$ another constant. Thus
$\dot{a}^2/a^4$ tends to a constant value. Therefore $a\sim t^{-1}$ near this limit, and this contradicts that $H\to 0$.  Another possible dangerous limit is $H\to 0$, $a\to 0$ and $f'(\phi)\to \infty$ in such a way that $H^2 f'(\phi)^2/a^2\to 0$, since for this limit $\dot{\phi}\to \infty$.  We suggest however that this limit do not take place.
In fact, the limit $H\to c$ implies near this region that $a(t)\sim \exp(ct)$, and $a(t)\neq 0$ even when $c\to 0$. 
 
 If the suggestion made above is true, then there should exist a global minimum $\dot{\phi}_1$ and a global maximum $\dot{\phi}_2$. Therefore
$$
\dot{\phi}_2\leq\dot{\phi}\leq\dot{\phi}_1,
$$
and, by simple integration of the last expression, it follows that
$$
 \phi_0+\dot{\phi}_1t\leq\phi\leq\phi_0+\dot{\phi}_2t,\qquad t\geq 0,
$$
\be\lb{bondol}
 \phi_0+\dot{\phi}_2 t\leq\phi\leq\phi_0+\dot{\phi}_1t,\qquad t\leq 0.
\ee
Again, the values of $\phi$ are bounded by two linear time functions, and therefore are bounded for any finite time $t$.

\section{The possibility of eternal universes for $k\neq 0$}

In the present section the possibility of having eternal solutions is considered, when the spatial curvature $k=\pm 1$ is turned on. However, the results obtained below are less general than the ones of the previous sections. In fact, the analysis when the spatial curvature $k$ is turned on 
is more difficult than the case $k=0$.

As before, it is assumed that $f'(\phi)$ is never zero and is never divergent for any finite value of $\phi$.
In other words, there are no vertical asymptotes at finite $\phi$ values.
Furthermore, we will be working in the branch for which $\dot{\phi}$ is bounded for any finite time.

The situation with $k=1$ will be analysed first. The case $H\to\pm\infty$ and $\ddot{\phi}\to\pm \infty$ with $a\to\infty$ is identical to the case with $k=0$, which was shown to be non singular. 
Consider now the case $H\to\pm\infty$ and $\ddot{\phi}\to\pm \infty$ with $a\to 0$ . The equation (\ref{ecu1}) shows that $1+H\dot{f}\to 0$ and thus $\dot{\phi}\to 0$.
By combining (\ref{ecu1}) with (\ref{ecu2}), together with the condition that $\dot{\phi}\to 0$ gives that
$$
2\bigg(\dot{H}-\frac{1}{a^2}\bigg)(1+\dot{f}H)+\bigg(H^2+\frac{1}{a^2}\bigg)(1+\ddot{f})\to 0.
$$
But since $1+\dot{f}H\to +0$ it follows that $(\dot{H}-a^{-2})^2>>H^4, a^{-4}$. Otherwise the first term would not compensate the second one.
This means that $\dot{H}^2>>H^4$ and $\dot{H}^2> a^{-4}$.
By taking this into account, it follows from (\ref{ecu3}) that
$$
\ddot{\phi}\sim 3 \bigg(H^2+\frac{1}{a^2}\bigg) f'(\phi)\dot{H}.
$$
The last equations combine to give
$$
2\bigg(\dot{H}-\frac{1}{a^2}\bigg)(1+\dot{f}H)+\bigg(H^2+\frac{1}{a^2}\bigg)^2f'(\phi)^2\dot{H}\to 0.
$$
But as $\dot{H}^2>>a^{-4}$, the last equation implies that
$$
2\dot{H}(1+\dot{f}H)+\bigg(H^2+\frac{1}{a^2}\bigg)^2f'(\phi)^2\dot{H}\to 0.
$$
However $1+\dot{f}H\to +0$, and the remaining terms multiplying $\dot{H}$ are strictly positive, thus this is never satisfied.
The same conclusions hold for $a$ taking any finite value $a_0$. No singularity will appear in this situation.

Consider now the possibility that $H^2<\infty$, $\ddot{\phi}\to\infty$ and $a\to 0$. If $H \to H_0$
then, near the singularity, $a\sim \exp(H_0 t)$ which contradicts our hypothesis that $a\to 0$. Thus, the only possibility is $H_0=0$.
Thus, as $H\to 0$ and $a\to 0$, a simple inspection shows that the equation (\ref{ecu1}) is never satisfied. If instead, one consider the same situation
but with $a\to\pm \infty$, this case reduce to the one with $k=0$ for which, as shown in previous sections, there are no singularities.

Consider now the possibility that $H^2<\infty$ and $a\to a_0$ and $\dot{H}\to \pm \infty$. Since $\dot{\phi}$ is bounded, equation (\ref{ecu1}) shows
that $1+\dot{f}H\geq 0$. The  equation (\ref{ecu2}) together with (\ref{ecu3}) gives the following requirement
$$
-2\dot{H}\bigg[1+\dot{f}H+\bigg(H^2+\frac{1}{a^2}\bigg)f'(\phi)^2\bigg]\to c,
$$
with $c$ a constant. But the term in parenthesis is strictly positive since $1+\dot{f}H>0$. But our assumption is that $\dot{H}\to\pm \infty$, so the requirement is impossible to satisfy.

Finally, one has to check the case $H\to H_0$, $\dot{H}\to\pm \infty$ and $a\to 0$. As we saw above, this means that $H_0=0$.
The equation (\ref{ecu1}) may be satisfied when $1+\dot{f}H\to 0$. But as $H\to 0$, one has that $1+\dot{f}H \to 1$, which is a contradiction.

The analysis given above is valid for $\ddot{\phi}\to\pm\infty$. It is impossible to have a singularity when $k=1$ in this case. However, the situation for finite $\ddot{\phi}$
is more difficult to analyse than for the case $k=0$. The reason is that the analysis made in  (\ref{ci7})-(\ref{ci12}) get much more complicated when the term $k/a^2$ is turned on. 
Thus, we have obtained no conclusions in this case. In addition, for $k=-1$, the term $H^2+k/a^2$ can be zero if a potential singularity takes place at $\dot{a}=1$ and $a\to 0$,
since $H^2+k/a^2\to 0$. This zero appears multiplying the factor (\ref{ecu1}) and  complicates the analysis of the singularity. We hope to overcome
these technical difficulties in a future.

\section{Discussion}

The results of the present work are the following. For a Gauss Bonnet model without cosmological constant and zero spatial curvature, if $0<f'(\phi)\leq c \exp(\frac{\sqrt{8}}{\sqrt{10}}\phi)$, and the scalar field is in some specific branch described in the text,  then exists a large class of solutions that are eternal. These conclusions were also obtained when the cosmological constant is turned on. It is important however to emphasize that if the scalar field is in  other branch, then the presented conclusions do not apply and in fact  singular solutions may appear. The appearance of this solutions do not contradict the well known Hawking singularity theorems, since it is not necessarily true that these theories may be considered as GR coupled with matter satisfying the strong energy conditions.

The analysis when the spatial curvature $k$ is turned on is more complicated. The problem is that for $k=0$ the resulting differential system only involves the Hubble constant $H$ and the scalar field, while for $k=\pm 1$ the system involves also the scale factor, and this complicates the analysis considerably. However, some partial results about the singularities were found, independently of the form of the coupling $f(\phi)$. These  results are collected in the proposition of section 6 in the text, and exclude under certain circumstances some singularities in the scale factor as $a(t)\sim c/(t-t_0)^\beta$ with $\beta>0$. This result is independent on the form of the coupling $f'(\phi)$. We hope to overcome some technical problems and to obtain results related to the case $k=\pm 1$  in a near future.

\section*{Acknowledgments}
J.O.M and O.P.S are supported by the CONICET.

\end{document}